\documentclass[preprint,floats,prl,aps,unsortedaddress,superscriptaddress]{revtex4-2}

\usepackage[dvips]{graphicx}
\usepackage{amsfonts}
\usepackage{amsmath}
\usepackage{amssymb}
\usepackage{exscale}
\usepackage{eufrak}
\usepackage{afterpage}
\usepackage{upgreek}
\usepackage{color}	
\usepackage{braket}	
\usepackage{multirow}	
\usepackage{epstopdf}	
\usepackage{setspace}
\usepackage{hyperref}

\usepackage[normalem]{ulem}    

\begin{document}

\title{Universal Fabrication of Two-Dimensional Electron Systems in Functional Oxides}

\author{Tobias~Chris~R\"odel}
\affiliation{CSNSM, Univ. Paris-Sud, CNRS/IN2P3, Universit\'e Paris-Saclay, 91405 Orsay, France}
\affiliation{Synchrotron SOLEIL, L'Orme des Merisiers, Saint-Aubin-BP48, 91192 Gif-sur-Yvette, France}
\author{Franck~Fortuna}
\affiliation{CSNSM, Univ. Paris-Sud, CNRS/IN2P3, Universit\'e Paris-Saclay, 91405 Orsay, France}
\author{Shamashis Sengupta}
\affiliation{Laboratoire de Physique des Solides, Univ. Paris-Sud, CNRS, Universit\'e Paris-Saclay, 
91405 Orsay, France}
\author{Emmanouil~Frantzeskakis}
\affiliation{CSNSM, Univ. Paris-Sud, CNRS/IN2P3, Universit\'e Paris-Saclay, 91405 Orsay, France}
\author{Patrick~Le~F\`evre}
\affiliation{Synchrotron SOLEIL, L'Orme des Merisiers, Saint-Aubin-BP48, 91192 Gif-sur-Yvette, France}
\author{Fran\c{c}ois~Bertran}
\affiliation{Synchrotron SOLEIL, L'Orme des Merisiers, Saint-Aubin-BP48, 91192 Gif-sur-Yvette, France}
\author{Bernard~Mercey}
\affiliation{CRISMAT, ENSICAEN-CNRS UMR6508, 6 bd. Mar\'echal Juin, 14050 Caen, France}
\author{Sylvia~Matzen}
\affiliation{Institut d'Electronique Fondamentale, Univ. Paris-Sud, CNRS, Universit\'e Paris-Saclay, 
91405 Orsay, France}
\author{Guillaume~Agnus}
\affiliation{Institut d'Electronique Fondamentale, Univ. Paris-Sud, CNRS, Universit\'e Paris-Saclay, 
91405 Orsay, France}
\author{Thomas~Maroutian}
\affiliation{Institut d'Electronique Fondamentale, Univ. Paris-Sud, CNRS, Universit\'e Paris-Saclay, 
91405 Orsay, France}
\author{Philippe~Lecoeur}
\affiliation{Institut d'Electronique Fondamentale, Univ. Paris-Sud, CNRS, Universit\'e Paris-Saclay, 
91405 Orsay, France}
\author{Andr\'es~Felipe~Santander-Syro}
\email{andres.santander-syro@universite-paris-saclay.fr}
\affiliation{CSNSM, Univ. Paris-Sud, CNRS/IN2P3, Universit\'e Paris-Saclay, 91405 Orsay, France}

\maketitle

This is the peer reviewed version of the following article:
T.~C.~R\"odel, F.~Fortuna, S.~Sengupta, E.~Frantzeskakis, P.~Le~F\`evre, F.~Bertran, B.~Mercey, 
S.~Matzen, G.~Agnus, T.~Maroutian, P.~Lecoeur, and A.~F.~Santander-Syro. 
\emph{Universal fabrication of 2D electron systems in functional oxides.}  
Adv. Mater. {\bf 28}, 1976-1980 (2016),
which has been published in final form at
\url{https://doi.org/10.1002/adma.201505021}. 
This article may be used for non-commercial purposes in accordance with Wiley Terms and Conditions 
for Use of Self-Archived Versions.

\pagebreak

\subsection*{Table of contents entry}
\textbf{Two-dimensional electron systems (2DESs) 
in functional oxides are promising for applications}, 
but their fabrication and use, essentially limited to SrTiO$_3$-based heterostructures, 
are hampered by the need of growing complex oxide over-layers 
thicker than 2~nm using evolved techniques. 
This work shows that thermal deposition of a monolayer of an elementary reducing agent 
suffices to create 2DESs in numerous oxides.

\textbf{Keyword}: 2DES in oxide surfaces and interfaces.\\

\begin{figure}[h]
    \begin{center}
        \includegraphics[clip, width=15cm]{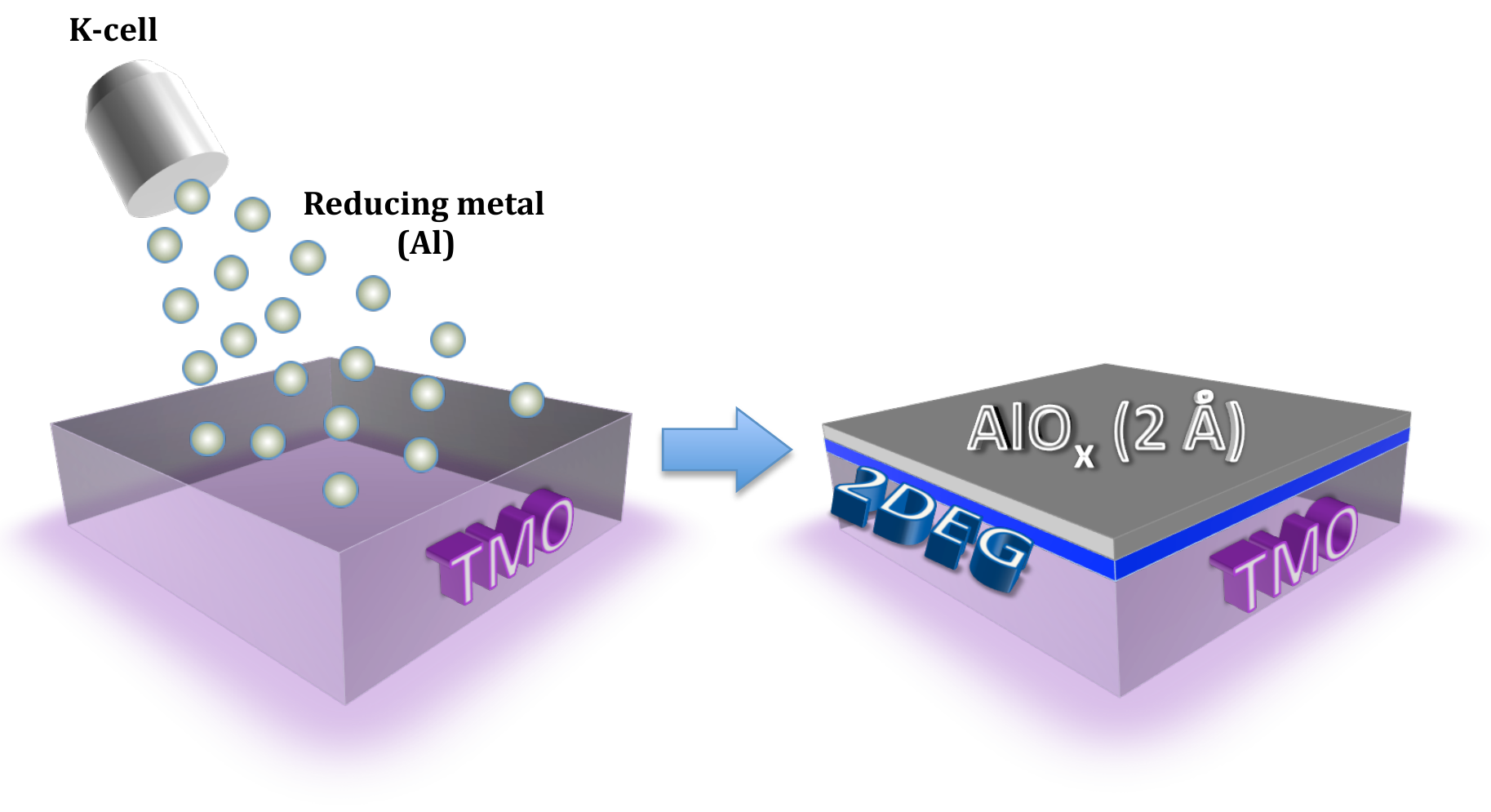}
    \end{center}
\end{figure}

\pagebreak

A critical challenge of modern materials science
is to tailor novel states of matter 
suitable for future applications beyond semiconductor technology. 
Two-dimensional electron systems (2DESs) 
in multi-functional oxides~\cite{Ohtomo2004} 
can show metal-to-insulator transitions~\cite{Thiel2006}, 
superconductivity~\cite{Reyren2007,Caviglia2008}, 
magnetism~\cite{Brinkman2007,Li2011,Bert2011}, 
or spin-polarized states~\cite{Caviglia2010,BenShalom2010,Santander-Syro2014},
and are thus an active field of current research~\cite{Takagi2010,Mannhart2010,Hwang2012}. 
However, the fabrication of 2DESs in oxide heterostructures, like LaAlO$_3$/SrTiO$_3$,  
requires growing a layer of binary (\emph{e.g.} Al$_2$O$_3$) 
or ternary (\emph{e.g.} LaAlO$_3$) oxides 
with a ``critical thickness'' of at least 20~\AA~using evolved deposition techniques, 
such as pulsed laser deposition~\cite{Ohtomo2004,Nakagawa2006,Takagi2010,Mannhart2010,
Hwang2012,Chen2011,Lee2012,Delahaye2012}.  
Thus, the reproducibility of their properties 
depends crucially on the growth parameters, while their fabrication 
is complex, expensive, and unsuitable for mass production.
Moreover, the existence of a critical thickness of 20~\AA~for the onset of conductivity 
severely limits the control of the 2DES’s properties, 
hampering tunneling spectroscopy studies or applications 
that rely on charge or spin injection~\cite{Lesne2014}.
Similarly, the realization of 2DESs at the surface of SrTiO$_3$ or other oxides 
requires the use of intense UV or X-ray synchrotron radiation, 
to desorb oxygen from the surface~\cite{Santander-Syro2011,Meevasana2011,
Santander-Syro2012,King2012,Bareille2014,Roedel2014,Walker2014,Roedel2015}.
Thus, these 2DESs can be only manipulated and studied in ultra-high vacuum (UHV), 
to preserve the vacancies from re-oxidation,
and are obviously not suited for experiments or applications at ambient conditions. 

Here we demonstrate a new,
wholly general, extremely simple and cost-effective method 
to generate 2DESs in functional oxides.
We use thermal evaporation from a Knudsen cell
to deposit, at room temperature in UHV, an atomically-thin layer 
of an elementary reducing agent, such as pure aluminum,
on the oxide surface.
Due to an efficient redox reaction, the Al film pumps oxygen from the substrate, 
oxidizes into insulating AlO$_x$, and forms a pristine, homogeneous 2DES
in the first atomic planes of the underlying oxide. 
The principle of redox reactions induced by metals
at the surface of oxides is well documented~\cite{Fu2005,Fu2007}.
However, the simple idea of using a pure, elementary reducing agent
to create a 2DES at a metal-oxide interface was not explored so far.
This overcomes the complexity of growing an oxide thin film, 
the requirement of a critical thickness of insulating capping layer
to create the 2DES in UHV,
and the necessity, in the case of surfaces, 
of strong synchrotron radiation to desorb oxygen.
As a novel application, we extend this method to generate
a 2D metallic state at the surface of the room-temperature ferroelectric BaTiO$_3$.
Such hitherto unobserved coexistence of ferroelectricity 
and 2D conductivity in the same material is promising for functional devices 
using ferroelectric resistive switching~\cite{Kim2013,Tra2013}.
This new, simpler and cheaper, fabrication route for 2DESs is thus adaptable 
to numerous oxides, given that oxygen vacancies are
shallow donors in these materials, resulting in itinerant
electrons.    
Moreover, this technique is scalable to industrial production, 
and ideally suited for applications that rely on charge or spin injection
and for the realization of mesoscopic devices.
  
\begin{figure*}
    \begin{center}
        \includegraphics[clip, width=16cm]{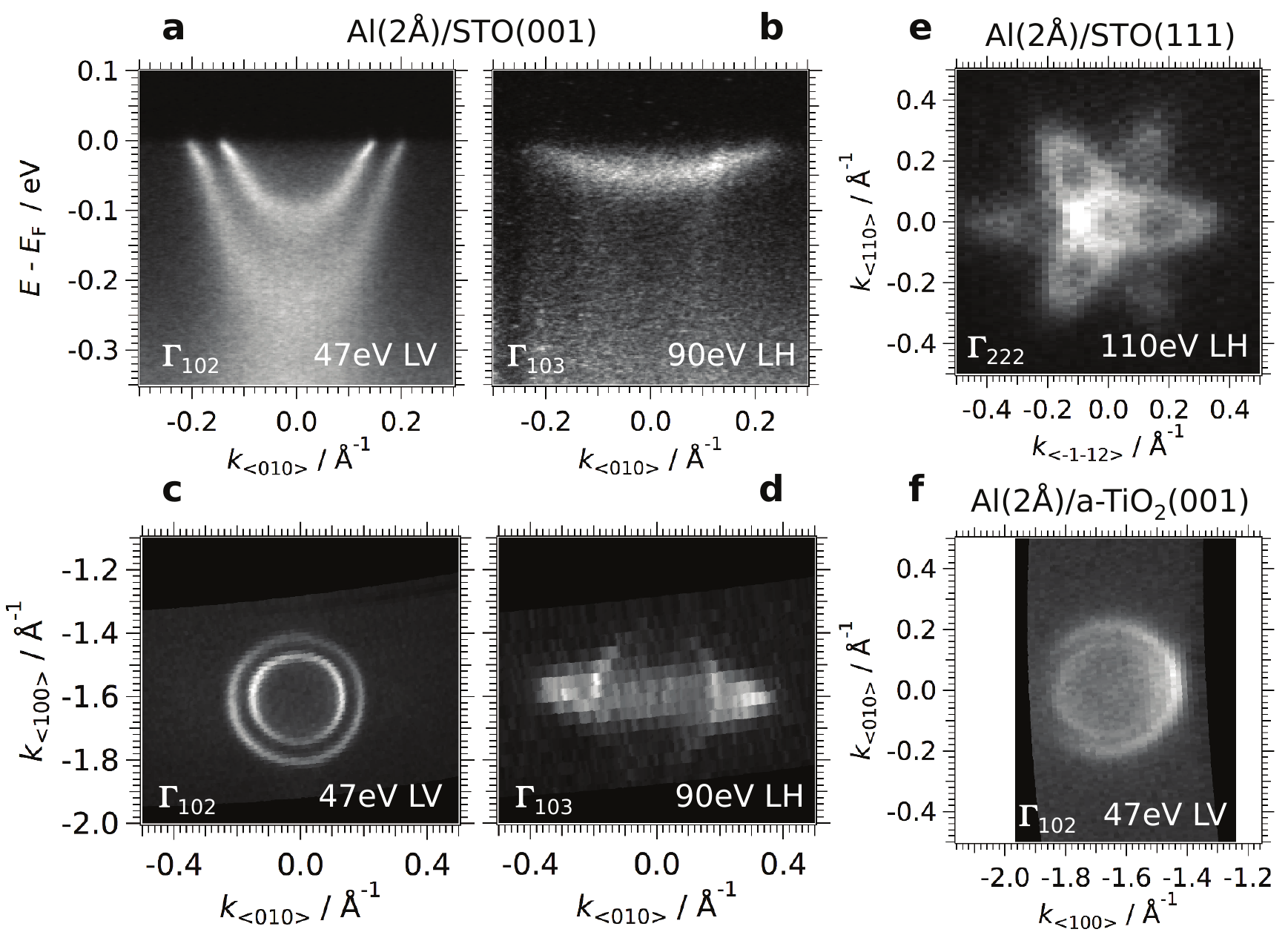}
    \end{center}
    \caption{\label{fig:STO001} \footnotesize{
        (a,~b) ARPES energy-momentum intensity maps measured 
        at the Al(2\AA)/SrTiO$_3$(001) interface prepared \emph{in-situ},
        using respectively $47$~eV linear vertical (LV) 
        and $90$~eV linear horizontal (LH) photons. 
        (c,~d) Corresponding Fermi surface maps. 
        Data at $h\nu = 47$~eV were measured around the $\Gamma_{102}$ point,
        while data at $h\nu = 90$~eV were measured around $\Gamma_{103}$.
        (e,~f) Fermi surface maps measured at the Al/SrTiO$_3$(111) 
        and Al/TiO$_2$(001) anatase interfaces prepared \emph{in-situ}.
        Unless specified otherwise, all spectra in this and remaining figures
        were measured at $T = 8$~K.
        }
      }     
\end{figure*}

The existence of a 2DES at the interface between the oxidized Al layer and SrTiO$_3$(001),
SrTiO$_3$(111) and anatase-TiO$_2$(001) 
is evidenced by our angle-resolved photoemission spectroscopy (ARPES) data 
presented in \textbf{Figure}~\ref{fig:STO001}
--see the \textcolor{blue}{Supporting Information} for details about the surface preparation,
Al deposition, and ARPES measurements. 
For simplicity, and to recall that we are simply depositing pure Al 
(\emph{not aluminum oxide}) on top of the oxide surfaces, 
all throughout this paper we note the resulting oxidized Al capping layer simply as ``Al",
specifying in parenthesis the evaporated thickness.
The energy-momentum and Fermi surface maps formed by the $t_{2g}$ orbitals,
shown in Figure~\ref{fig:STO001}, agree with previous ARPES studies  
at the reduced surface of these materials~\cite{Santander-Syro2011,Meevasana2011,Plumb2014,
Roedel2014,Walker2014,Roedel2015},
demonstrating that in both cases the same 2DESs are observed.

Note that, instead of the local creation of oxygen vacancies using an intense UV beam, 
the evaporated Al reduces the whole surface homogeneously.
As a consequence, the data quality, evidenced by the line widths, 
is also much better than in previous studies. 
Thus, as shown in Figure~\ref{fig:STO001}(a), a kink and change in intensity 
in the dispersion of the light bands at $E \approx -30$~meV,
attributed to electron-phonon coupling~\cite{King2014},
can be very clearly distinguished. 

The Fermi-surface areas and, hence, the charge carrier densities
of the 2DESs at the Al/SrTiO$_3$(111) and Al/TiO$_2$(001) interfaces
are about $1.3$ and $2$ times larger than their counterparts 
at the surfaces reduced by photons,
probably due to a higher and more homogeneous concentration of oxygen vacancies. 

\begin{figure}
    \begin{center}
        \includegraphics[clip, width=8.5cm]{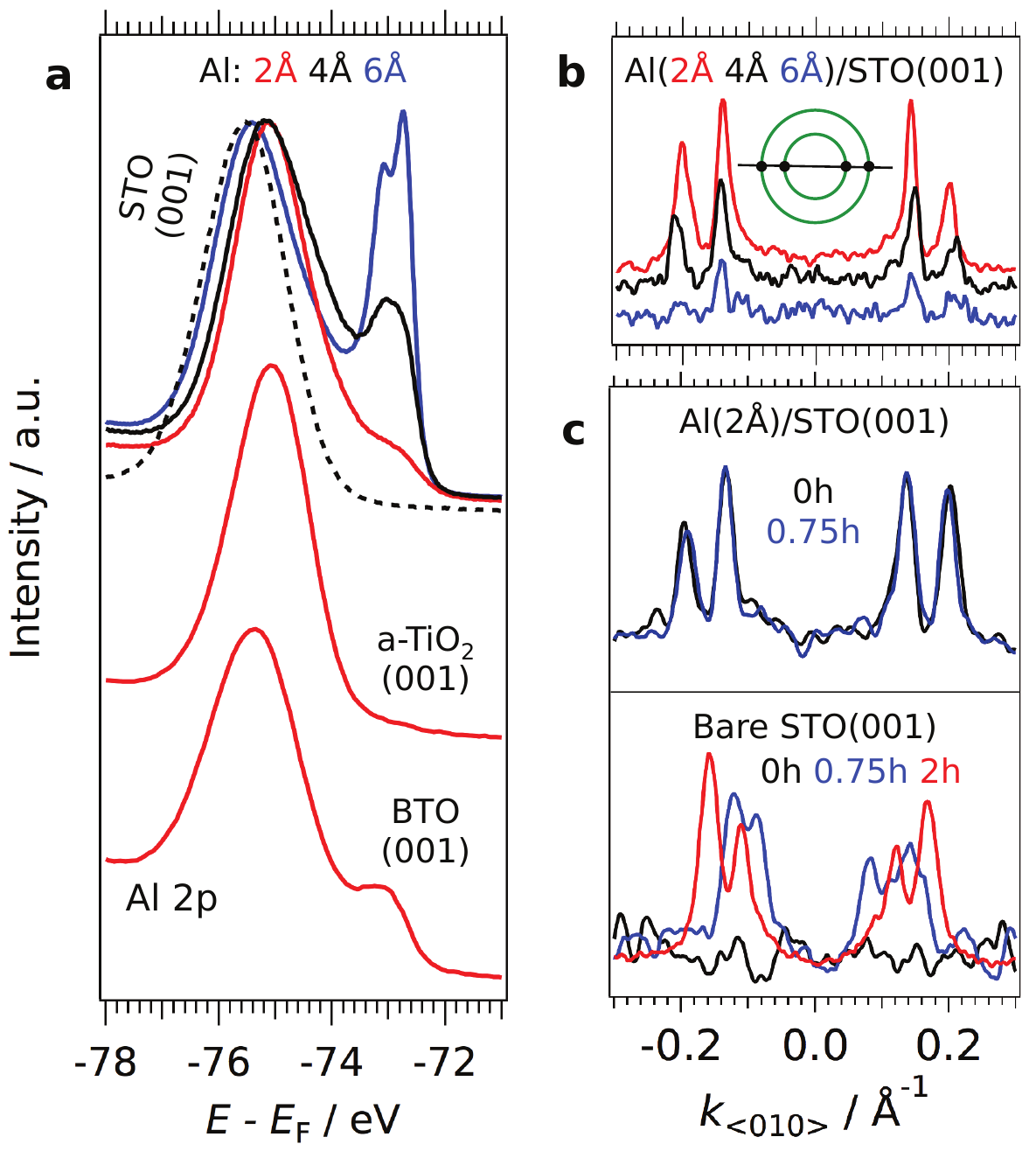}
    \end{center}
    \caption{\label{fig:Al2p} \footnotesize{
        (a) Angle-integrated spectra of the Al 2p peak of the Al/SrTiO$_3$(001)
        Al/TiO$_2$ anatase and Al/BaTiO$_3$(001) interfaces 
        measured at a photon energy of $h\nu=100$~eV. 
        The curves in different colors correspond to different thicknesses 
        of Al (red 2\AA, black 4\AA, blue 6\AA).
        The peak shape of the Al 2p peak indicates if the Al layer is oxidized or metallic.
        The dashed black curve corresponds to a fully oxidized Al layer, 
        obtained after annealing the sample with $4$~\AA~Al capping at 250$^{\circ}$C in UHV.
        (b) Momentum distribution curves (MDCs) at $E_F$, 
        along the Fermi-surface cut schematized in the inset,
        measured at the Al/SrTiO$_3$(001) interface at $hv=47$~eV 
        for different Al thicknesses. Peaks in the MDCs correspond to the Fermi momenta, 
        where the MDC cuts the Fermi surface. 
        The decrease in intensity of the MDCs for increasing Al thickness 
		is merely due to increasing damping of the photoemission signal.
		(c) MDCs integrated over $E_F \pm 5$~meV for increasing UV exposure times 
		on the Al/SrTiO$_3$(001) interface and the bare STO surface 
		measured under identical conditions. 
		The similarity of the two MDCs at the interface is in strong contrast 
		to the evolution under light irradiation of the MDCs at the bare surface. 
        }
      }     
\end{figure}

To understand the redox reaction at the Al/oxide interface, 
we probed the oxidation state of Al by measuring the Al-$2p$ core levels, 
whose binding energies are very different for metallic and oxidized Al. 
As shown in \textbf{Figure}~\ref{fig:Al2p}(a), the two contributions can be distinguished
in the Al(6\AA)/STO spectrum (blue curve), with the metallic component 
around $72.5$~eV binding energy and the oxidized part around $75$~eV binding energy. 
In contrast, the metallic Al component decreases for a thinner 4\AA~film (black curve),
and the deposition of only 2\AA~of pure Al results in a nearly fully oxidized 
film of Al (red curves). 
In other words, \emph{an ultra-thin layer of 2\AA~of pure Al is sufficient 
to pump the oxygen from the surface region of all the oxides studied in this work}.
The spatial distribution of the oxygen vacancies close to the interface 
is discussed in the Supporting Information.
Note that the oxidation of the metallic Al results in an increased layer thickness: 
as the mass density of Al is $2.7$~g/cm$^3$ and the one of amorphous Al$_2$O$_3$ 
is about $4$~g/cm$^3$, the deposition of 2\AA~of Al 
yields an oxidized Al film of 2.5\AA.

To determine if the thickness of the Al-layer has an influence 
on the electron density of the 2DES, we turn to the momentum distribution curves (MDCs) 
at the Fermi level as shown in Figure~\ref{fig:Al2p}(b). 
As can be seen in Figure~\ref{fig:Al2p}(b), the Fermi momenta 
are essentially the same, within $0.01$~\AA$^{-1}$ 
for the 2\AA~(red MDC), 4\AA~(black) and 6\AA~(blue) thick Al films. 
As the 2D density of electrons depends solely on the Fermi momenta,
it is clear that this electron density already saturates at an Al film thickness of 2\AA.
Thus, our method overcomes the necessity of a ``critical thickness'' of capping layer
to generate a 2DES in UHV.

Previous studies on the bare surface of SrTiO$_3$ prepared \textit{in-situ} 
showed that synchrotron UV-irradiation was necessary to create the oxygen vacancies
responsible for the 2DES~\cite{Meevasana2011,Plumb2014,Walker2014,Roedel2015}. 
This is again demonstrated in the lower panel of Figure~\ref{fig:Al2p}(c),  
which shows the evolution with time of the MDC at $E_{F}$ 
upon UV-irradiation on a bare SrTiO$_3$(001) surface. 
While the 2DES is absent at $t=0$h (black MDC), 
its carrier density increases up to saturation upon UV irradiation (blue and red MDCs), 
as denoted by the increase of $k_F$ for increasing exposure times.

Contrary to the bare surface, there is \emph{no measurable influence} 
of the UV irradiation on the electronic structure of the Al/SrTiO$_3$ system, 
neither on the charge carrier density nor on the line-shapes or spectral weight of the 2DES, 
as demonstrated in the top panel of Figure~\ref{fig:Al2p}(c): 
the MDCs at $E_F$ show a stable subband structure and a maximum electron density 
from the very beginning of the measurements. 
This indicates that the oxygen vacancy concentration and distribution, 
due the redox reaction at the interface between Al and SrTiO$_3$, 
is already saturated and stable upon irradiation.

As demonstrated in the \textcolor{blue}{Supporting Information}, 
the 2DES at the interface of oxidized Al/SrTiO$_3$ is stable also at room temperature,
while the deposition of an Al film of 10\AA~or more on SrTiO$_3$ 
minimizes the re-oxidation of vacancies in air. 

\begin{figure*}
    \begin{center}
        \includegraphics[clip, width=16cm]{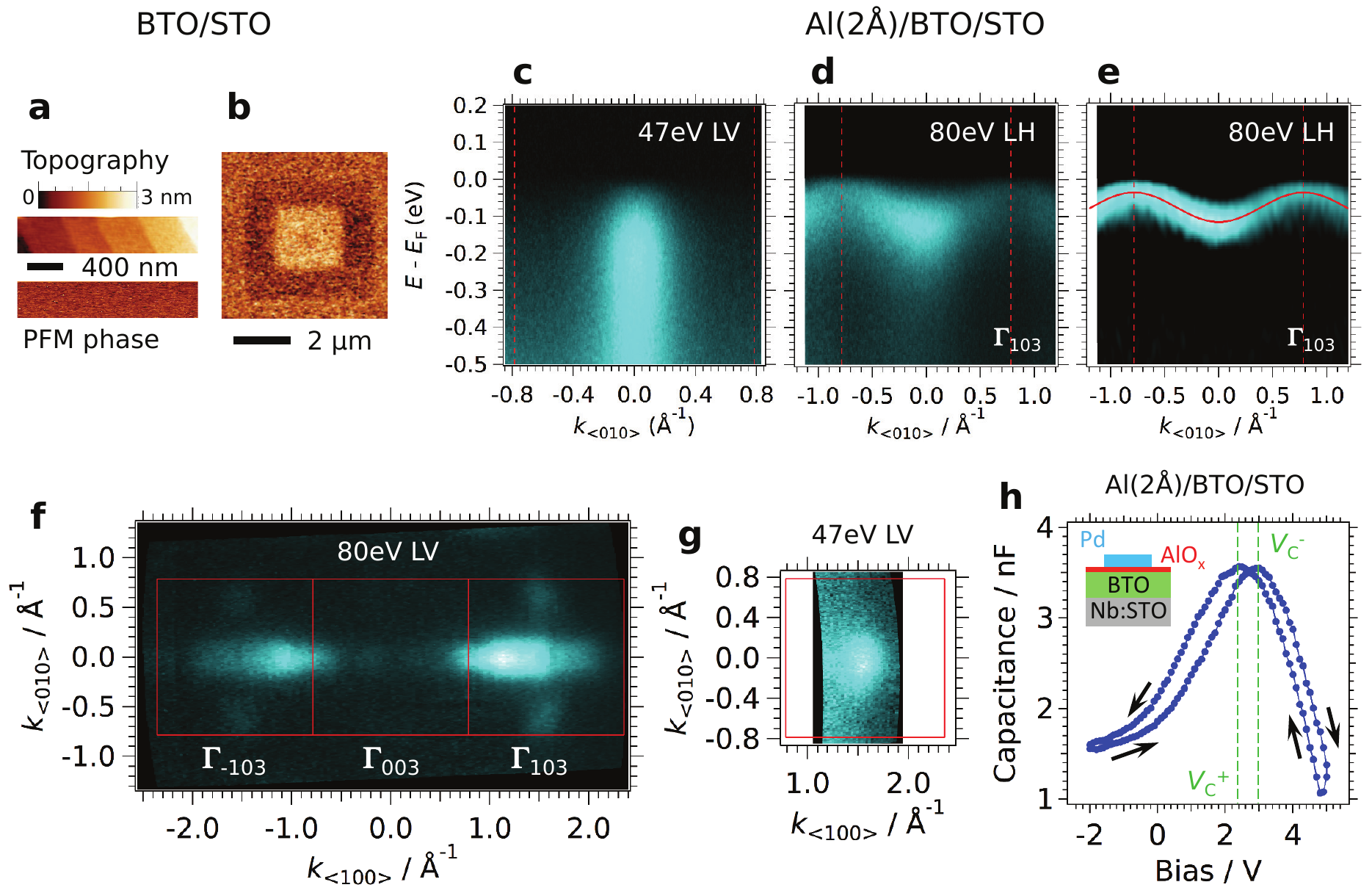}
    \end{center}
    \caption{\label{fig:BTO001} \footnotesize{
        (a)~AFM topography and corresponding PFM phase signal
        measured on a $30$~nm-thick BaTiO$_3/$Nb:SrTiO$_3$ film.
        No ferroelectric domains could be detected in the as-grown film, 
        while such domains can be written, as shown in (b), 
        with $+6$~V on the AFM tip in the outer square ($4 \times 4$~$\mu$m$^2$) 
        and $-6$~V in the inner square ($2 \times 2$~$\mu$m$^2$). 
        (c,~d) ARPES energy-momentum intensity maps 
        at the Al(2\AA)/BaTiO$_3$ interface prepared \emph{in-situ},
        using respectively $47$~eV LV and $80$~eV LH photons,
        the latter being close to $\Gamma_{103}$.
        (e)~Second energy-derivative (negative values) of the ARPES map in (d).
        Vertical dashed red lines in (c)-(e) are the Brillouin-zone edges.
        The red curve in (e) is a cosine fit to the heavy band. 
        (f,~g)~Spectral weight integrated over $E_F \pm 30$~meV
        at the Al(2\AA)/BaTiO$_3$ interface 
        using $80$~eV LV and $47$~eV LV photons, respectively. 
        (h)~Capacitance-voltage curve on the Al(2\AA)/BaTiO$_3$ interface 
        measured previously by ARPES, showing the butterfly shape 
        characteristic of a ferroelectric hysteresis. 
        A Pd circular pad and the Nb:STO substrate were used 
        as top and bottom electrodes, respectively. 
        Note that, due to the voltage drop through the thin alumina layer, 
        the voltages $V_{c}^{+}$ and $V_{c}^{-}$ required to reverse the polarization
        are rather high. Thus, it was not possible to perform a polarization reversal in PFM mode.
        }
      }     
\end{figure*}
We now show that the deposition of an ultra-thin Al film 
can also be used to create a 2DES at the surface 
of the room-temperature ferroelectric BaTiO$_3$ (BTO),
thus constituting a new type of confined metallic state 
on a truly room-temperature functional oxide.
Our BTO samples are (001)-oriented thin films (thickness $30$~nm) 
epitaxially grown on SrTiO$_3$(001) 
--see \textcolor{blue}{Supporting Information} for details 
about thin-film growth, piezo-response force microscopy (PFM), 
and capacitance-voltage (C-V) measurements. 
In contrast to the bulk crystals, which 
usually exhibit ferroelectric-domain stripes of period $\sim 50 - 200$~nm, 
even down to $100$~nm scale thicknesses~\cite{Schilling2006},
the $30$~nm-thick BTO films deposited on Nb:STO show a single domain state,
with the ferroelectric polarization aligned along the $[001]$ axis, due to the in-plane 
compressive strain induced by the epitaxial growth~\cite{Pertsev1998,Choi2004,Chen2013}.

The absence of ferroelectric domains and the local reversibility 
of the polarization are demonstrated in \textbf{Figures}~\ref{fig:BTO001}(a,~b)
by the simultaneous atomic force microscopy (AFM) and PFM images 
of a BaTiO$_3/$Nb:SrTiO$_3$ thin film. 

The energy-momentum ARPES intensity maps of Figures~\ref{fig:BTO001}(c-e)
prove the formation of metallic itinerant states at the surface of the BTO(001) thin-film 
after deposition of 2~\AA~of Al.
The resulting 2DES is constituted of a light ($d_{xy}$-like) and a heavy ($d_{xz/yz}$-like)
electron pocket around $\Gamma$, best observed in Figures~\ref{fig:BTO001}(c,~d) 
with LV and LH polarizations, respectively.
Such light-polarization-dependent selection rules are typical for $t_{2g}$-like 
states observed at the surface
of other titanates, such as SrTiO$_3$ and anatase~\cite{Santander-Syro2011,Roedel2015}.
In the case of BaTiO$_3$, the light band forms a strong peak of spectral weight
whose intensity is cut-off at $E_F$ 
--see Figure~\ref{fig:BTO001}(c).
Although we cannot observe a dispersive feature within this peak of intensity,
its binding energy indicates that the conduction band bottom 
is filled up with itinerant electrons.
The heavy band, on the other hand, presents a clear dispersion 
--see Figures~\ref{fig:BTO001}(d,~e). 
A tight-binding fit of this band, red curve in Figure~\ref{fig:BTO001}(e),
yields a band bottom of $-115$~meV at $\Gamma$, a band top of $-35$~meV at the zone edge, 
and an effective mass near $\Gamma$ of approximately $12 m_e$.

Figures~\ref{fig:BTO001}(f,~g) show that the spectral weight at $E_F$ 
is composed of a central disc formed by the light electron pocket,
best seen in Figure~\ref{fig:BTO001}(g), 
and two orthogonal Fermi-surface strips spanning the entire Brillouin zone,
formed by the heavy bands. 
These last correspond to the elliptical Fermi sheets observed at the surface
of SrTiO$_3$(001), as in Figure~\ref{fig:STO001}(d), but in the case
of BaTiO$_3$(001) they extend beyond the zone boundary, 
thus forming open Fermi sheets.
From Figure~\ref{fig:BTO001}(g), the distribution of spectral weight at $E_F$
for the circular Fermi surface spans a Fermi momentum 
$k_F \approx 0.15 \pm 0.02$~\AA$^{-1}$. 
The Fermi strips can be approximated as rectangles 
of long and short sides $k_{l} = 2 \times \pi / a$ 
(with $a = 4$~\AA~the size of the square unit cell at the BTO surface) 
and $k_{s} = 0.15 \pm 0.02$~\AA$^{-1}$.
From the total area $A_F$ enclosed by all the Fermi surfaces, 
the density of carriers of the 2DES at the BaTiO$_3$(001) surface is 
$n_{2D}^{\text{BTO(001)}} = A_F/(2\pi^2) 
\approx (2.8 \pm 0.4) \times 10^{14}$~cm$^{-2}$,
which is comparable to the density of states at the SrTiO$_3$
or anatase-TiO$_2$ surfaces.
The \textcolor{blue}{Supporting Information} presents additional data
showing the Fermi momenta extracted from fits to the spectra, 
and the photon-energy dependence of the electronic structure.

Finally, Figure~\ref{fig:BTO001}(h) shows a measurement of the capacitance-voltage curve
on the \emph{same} Al(2\AA)/BaTiO$_3$ interface that was measured by ARPES.
The ``butterfly'' shape, with a difference of about $0.5$~V 
between the two coercive voltages,
demonstrates that the BTO film is still ferroelectric after deposition of the Al layer
and ARPES measurements, thus keeping its functional behavior.

A 2DES at the surface of BaTiO$_3$ is in essence an intrinsic metal/ferroelectric interface.
Polarization switching of the bulk material, for instance by strain, 
could allow a direct gating of the 2DES,
while a sufficiently thick capping alumina layer protects it 
against re-oxidation at ambient conditions, 
and can be even used to draw metallic nano-circuits of intrinsic ferroelectric tunnel junctions.
Thus, this system provides a realistic platform for the realization of non-volatile memories
using ferroelectric resistive switching~\cite{Kim2013,Tra2013}
or for ultra-sensitive strain or pressure detectors.

In conclusion, the method we present here for realizing 2DESs in oxides 
has the advantages of simplicity and versatility 
--for instance, it can be readily implemented in many UHV setups,
allowing future investigations of 2DESs
in complex oxides using non-synchrotron based spectroscopic techniques, 
like tunneling or Raman spectroscopies. 
This method is also pertinent for the study of transport phenomena 
in mesoscopic oxide devices. 
Indeed, STO has emerged as an exciting nano-electronics device platform~\cite{Goswami2015},
owing to the existence of superconductivity, spin-orbit interaction and magnetism 
which are controllable with a gate voltage. 
Our work opens up new possibilities to explore these questions 
by making a class range of transition-metal oxide 2DESs suitable for transport,
including the surfaces which are candidates for hosting 
topological electronic states~\cite{Bareille2014,Roedel2014}.
Furthermore, the stability of the 2DES in ambient conditions
can be achieved through a sufficiently thick layer of oxidized Al.
This opens the possibility to integrate TMO 2DESs
into functional devices without the need of evolved deposition techniques.

\subsection*{Experimental Section}
The ARPES measurements were conducted at the CASSIOPEE beamline 
of Synchrotron SOLEIL (France). 
We used linearly polarized photons in the energy range $30-110$~eV and 
a hemispherical electron analyzer with vertical slits.
The angular and energy resolutions were $0.25^{\circ}$ and 15~meV. 
The mean diameter of the incident photon beam was smaller than 100~$\upmu$m.
The samples were cooled down to $T=8$~K before measuring.
Unless specified otherwise, all data were taken at that temperature.
The results have been reproduced on more than 10 samples for SrTiO$_3$(001), 
and on at least two samples for other surface orientations and for TiO$_2$ anatase,
and on 3 thin films of BaTiO$3$/SrTiO$_3$(001).
All through this paper, reciprocal-space directions $\langle hkl \rangle$ and planes $(hkl)$ 
are defined in the conventional cell of each material (cubic for SrTiO$_3$, 
simple tetragonal for anatase and BaTiO$_3$).
The indices $h$, $k$, and $l$ of $\Gamma_{hkl}$ correspond to
the reciprocal lattice vectors of the cubic unit cell (SrTiO$_3$ and BaTiO$_3$) 
or body-centered tetragonal unit cell (anatase).

Additional details on the sample and surface preparation, the Al deposition conditions,
and the piezo-response force microscopy and capacitance-voltage measurements
can be found in the \textcolor{blue}{Supporting Information}.

\subsection*{Acknowledgements}
We thank illuminating discussions with M.~Gabay and M.~J.~Rozenberg,
and V. Pillard for help with the sample preparation.
This work is supported by public grants 
from the French National Research Agency (ANR), project LACUNES No ANR-13-BS04-0006-01, 
and the ``Laboratoire d'Excellence Physique Atomes Lumi\`ere Mati\`ere'' 
(LabEx PALM project ELECTROX) overseen by the ANR as part of the 
``Investissements d'Avenir'' program (reference: ANR-10-LABX-0039).
T.~C.~R. acknowledges funding from the RTRA--Triangle de la Physique (project PEGASOS).
A.F.S.-S. thanks support from the Institut Universitaire de France.

\section*{Supporting Information}

\subsection*{Sample and surface preparation and Al deposition}
The non-doped, polished crystals of SrTiO$_3$ were supplied by CrysTec~GmbH, 
the anatase crystals by SurfaceNet GmbH.
To prepare the SrTiO$_3$ surfaces, the samples were ultrasonically agitated 
in deionized water, subsequently etched in buffered HF and annealed at 950$^{\circ}$C 
for three hours in oxygen flow. This results in Ti-terminated (001) or (111) 
surfaces with terraces of width $50$~to~$200$~nm separated by steps, 
as verified by atomic-force microscopy (AFM) in contact mode (data not shown here).

The BTO 300\AA-thick films were prepared by Laser-MBE using a sintered BTO target. 
A Kr-F excimer laser was used for the deposition.
The substrate, which was etched prior to the deposition to obtain a TiO$_2$ terminating layer, 
was glued to the heater with silver paste. 
The growth of the film, monitored by reflection high-energy electron diffraction (RHEED), 
was carried out at 650 $^{\circ}$C in 5x10$^{-4}$ mbar oxygen pressure with 0.1 \% ozone. 
RHEED oscillations were used to measure the deposited thickness. 
At the end of the deposition, the films were cooled down in 6x10$^{-3}$ mbar oxygen pressure,
always with 0.1\% ozone.

To clean the surfaces in UHV, the SrTiO$_3$ samples were annealed at a temperature 
$T=550-650$~$^\circ$C for $t=10-90$~min at pressures lower than $p < 2\times10^{-8}$~mbar. 
The anatase crystals were prepared by Ar$^{+}$ sputtering ($U=1$~kV, $t=10$~min) 
and annealing cycles ($T=550-600$~$^\circ$C, $t=30$~min,) 
similar to the procedure described by Setvin~\textit{et al.}~\cite{Setvin2014a}. 
The surface of the BaTiO$_3$ thin films was cleaned by annealing the samples 
at temperatures $T = 500-550^{\circ}$~C for $5 - 30$~min. 
One of the samples was Ar$^+$ sputtered ($U = 500$~V, $t = 10$~min) prior to the UHV annealing,
without noticeable changes in the ARPES data.

The surface quality and the possible existence of surface reconstructions 
was probed by low-energy electron diffraction (data not shown here). 
The SrTiO$_3$(001) and BaTiO$_3$(001) surfaces are unreconstructed, 
whereas the (111) surface shows a $3 \times 3$ reconstruction,
and the anatase (001) surface shows a two-domain $1\times4$ reconstruction.

To create a local, high concentration of oxygen vacancies in the surface region 
of SrTiO$_3$, TiO$_2$ anatase, or BaTiO$3$, amorphous Al-films with thicknesses 
between $d=2-10$~\AA~were deposited on the prepared surface of the crystals. 
Aluminium was evaporated from a Knudsen cell using an alumina crucible. 
The growth rate was approximately $0.3$~\AA~/min, corresponding to a temperature of 
about 925$^{\circ}$C of the crucible. The Al-flux was calibrated 
prior to the evaporation using a quartz microbalance. The cleanliness of the
deposit was checked by evaporating a thin Al-film on a Cu substrate 
where no oxidation could be detected by Auger spectroscopy. 
The temperature of the crystals ranged between $T=25-100$~$^\circ$C 
during the Al deposition.

\subsection*{Piezo-response force microscopy and capacitance-voltage measurements}
For the PFM measurements, a probing signal of $2$~V$_{\textrm{pp}}$ 
at a frequency of 25~kHz was applied to a Co/Cr coated cantilever 
with $\sim 5$~N/m force constant. A lower probing signal of $0.5$~V$_{\textrm{pp}}$ 
was also used with no change on the observed phase images 
such as the ones shown in Figure~3(a) of the main text.
In order to assess the ferroelectric character of the BTO film measured by ARPES, 
$300$~$\mu$m-diameter Pd electrodes ($200$~nm thickness) were deposited 
through a shadow mask on top of the Al oxide (AlO$_x$) layer. 
The C-V measurements were performed using a LCR meter with a $30$~mV$_{\textrm{pp}}$ 
AC amplitude at $10$~kHz, while a source-meter allowed for the DC biasing with $0.1$~V steps 
of $500$~ms duration.
The C-V curve, Figure~3(h) of the main text, shows the characteristic butterfly shape 
of a ferroelectric material. 
Note that due to the ultra-thin AlO$_x$ layer, 
the required voltages to reverse the polarization ($V_c$) are rather high 
and shifted towards the positive voltage side, 
indicating an internal upward electric field in the BTO layer. 
For these reasons it was not possible to achieve a polarization reversal 
in PFM configuration on this sample.

\subsection*{Homogeneity of the 2DES}
\begin{figure}
    \begin{center}
        \includegraphics[clip, width=5cm]{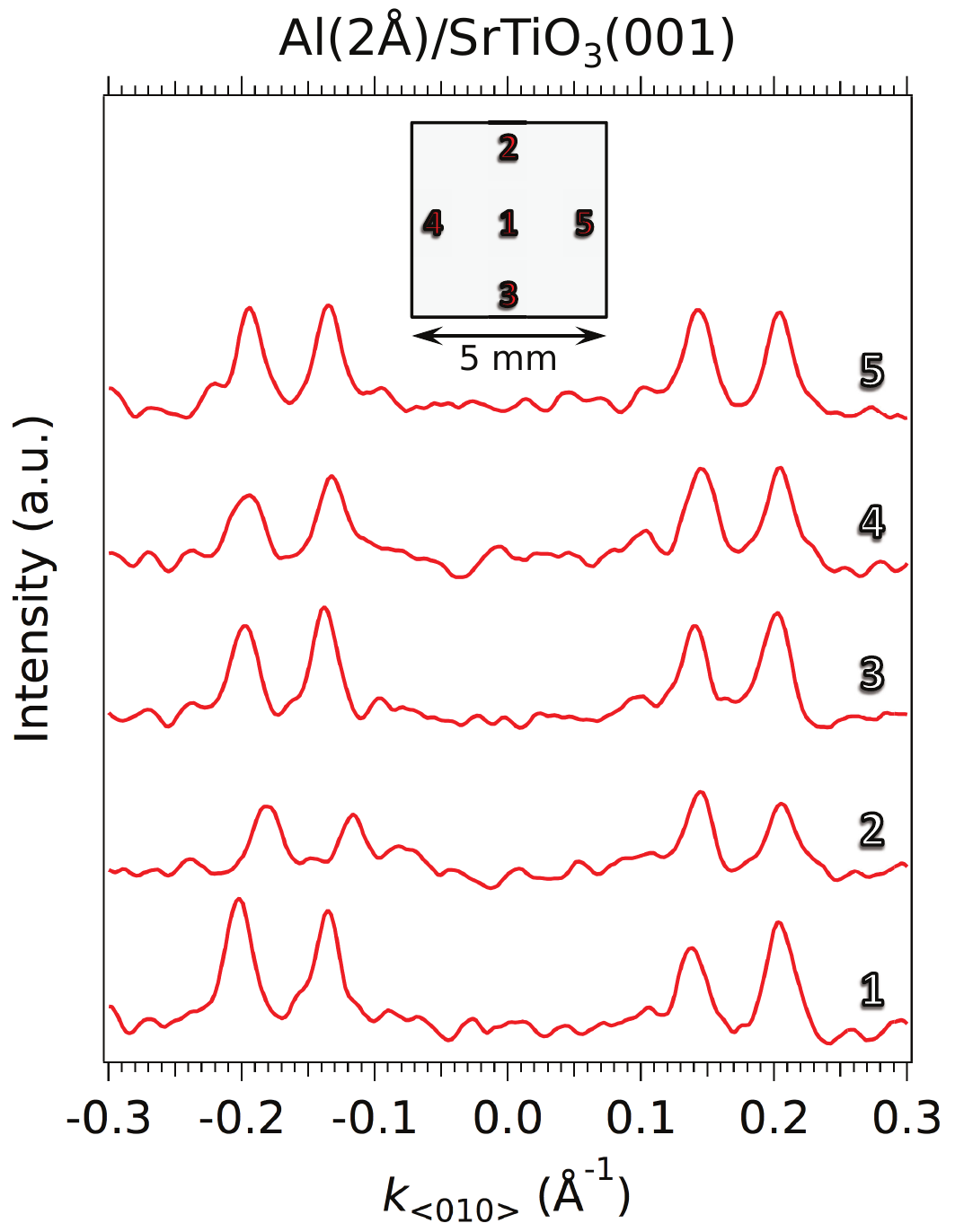}
    \end{center}
    \caption{\label{fig:STO001_homo} \footnotesize{
        (a) Momentum distribution curves integrated over $E_F \pm 5$~meV
        for the Al/SrTiO$_3$(001) interface prepared \emph{in-situ},
        measured at different positions (see inset) separated by at least 
        2~mm from each other -or over 100 times the size of the UV spot. 
        Each spectrum was obtained within minutes on a part of the sample 
        that had not been illuminated before.
        The Fermi momenta, given by the MDC peak positions, 
        are independent of the position in the sample, 
        demonstrating the homogeneity of the 2DES
        at the Al/SrTiO$_3$(001).
        }
      }     
\end{figure}

Previous studies of the 2DES at the surface of SrTiO$_3$ 
were conducted on fractured~\cite{Santander-Syro2011, Meevasana2011} 
or \emph{in-situ} prepared surfaces~\cite{Plumb2014, Roedel2014}. 
The fracturing process results in locally ordered surfaces~\cite{Guisinger2009}, 
while the \emph{in-situ} preparation results in an ordered surface. 
The fracturing process or intense UV light irradiation at low temperature 
(spot size $\approx 100\times100\upmu$m) create a local, high concentration of oxygen vacancies 
in the surface region of SrTiO$_3$ whose electrons (partly) dope the 2DES. 

In contrast to such spatial inhomogeneity of the 2DES 
in fractured or UV irradiated bare SrTiO$_3$, 
the fact that we cannot observe any changes induced by the synchrotron light 
at the Al/SrTiO$_3$ interface (Figure~2(c) of the main text) 
suggests that the underlying SrTiO$_3$ surface is reduced homogeneously.
This is explicitly shown in \textbf{Figure~}\ref{fig:STO001_homo}, 
which presents the momentum distribution curves at $E_F$ 
measured at five different positions on the Al(2\AA)/SrTiO$_3$ interface. 
We observe that the Fermi momenta, given by the MDC peaks, are independent of the measurement point,
demonstrating the homogeneity of the interfacial 2DES over distances of several millimeters.

\subsection*{Temperature dependence of the electronic structure of the 2DES
			at the Al/SrTiO$_3$(001) interface}
\begin{figure*}
    \begin{center}
        \includegraphics[clip, width=16cm]{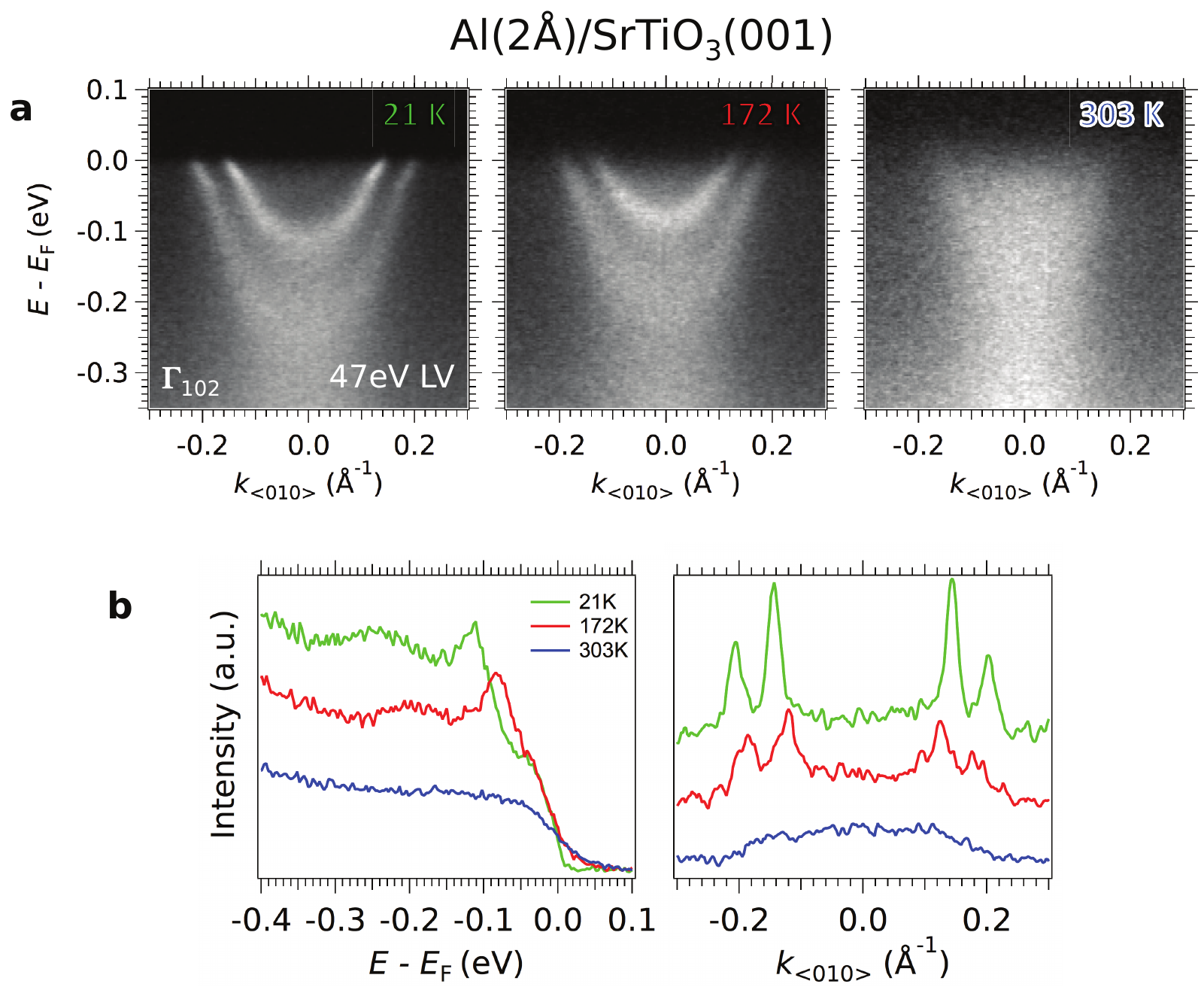}
    \end{center}
    \caption{\label{fig:STO001_temp} \footnotesize{
        (a) Energy-momentum map measured at the Al(2~\AA)/SrTiO$_3$(001) 
        interface prepared \emph{in-situ} at different temperatures 
        $T=$21~K, 172~K and 303~K. Data were collected around the $\Gamma_{102}$ point
        using LV photons at $h\nu = 47$~eV.
        (b) Energy and momentum distribution curves along $k_{<010>}=0$ 
        and the Fermi level of the $E-k$ maps in (a).
        }
      }     
\end{figure*}

\textbf{Figure~}\ref{fig:STO001_temp}(a) shows the energy-momentum maps at the Al/SrTiO$_3$ interface 
measured respectively at 21~K, 172~K, and 303~K, 
under the same conditions as the $E-k$ map in Figure~1(a) of the main text.
The dispersions of the two light bands of $d_{xy}$-character 
are still clearly visible at $T=172$~K although the line widths are increased 
due to thermal broadening. 
At $T=300$~K, the line widths are too large to identify the individual bands,
although the left branch of the outer band 
is still visible close to the Fermi level $E_F$.
Nevertheless, the spectral weight at the Fermi level 
demonstrates the existence and stability of the 2DES at room temperature. 
To compare the energy-momentum maps more directly, Figure~\ref{fig:STO001_temp}(b) 
shows the energy distribution curves at $k_{<010>}=0$ 
and the momentum distribution curves at Fermi level $E_F$. 
As can be seen from the value of the Fermi momenta $k_F$ of the peaks in the MDCs 
as well as the binding energies of the peaks in the energy distribution curves (EDCs), 
the charge carrier density decreases slightly when temperature increases.
We checked (not shown) that these results are reproducible upon thermal cycling.

\subsection*{Effect of exposure to ambient conditions}
\begin{figure*}
    \begin{center}
        \includegraphics[clip, width=16cm]{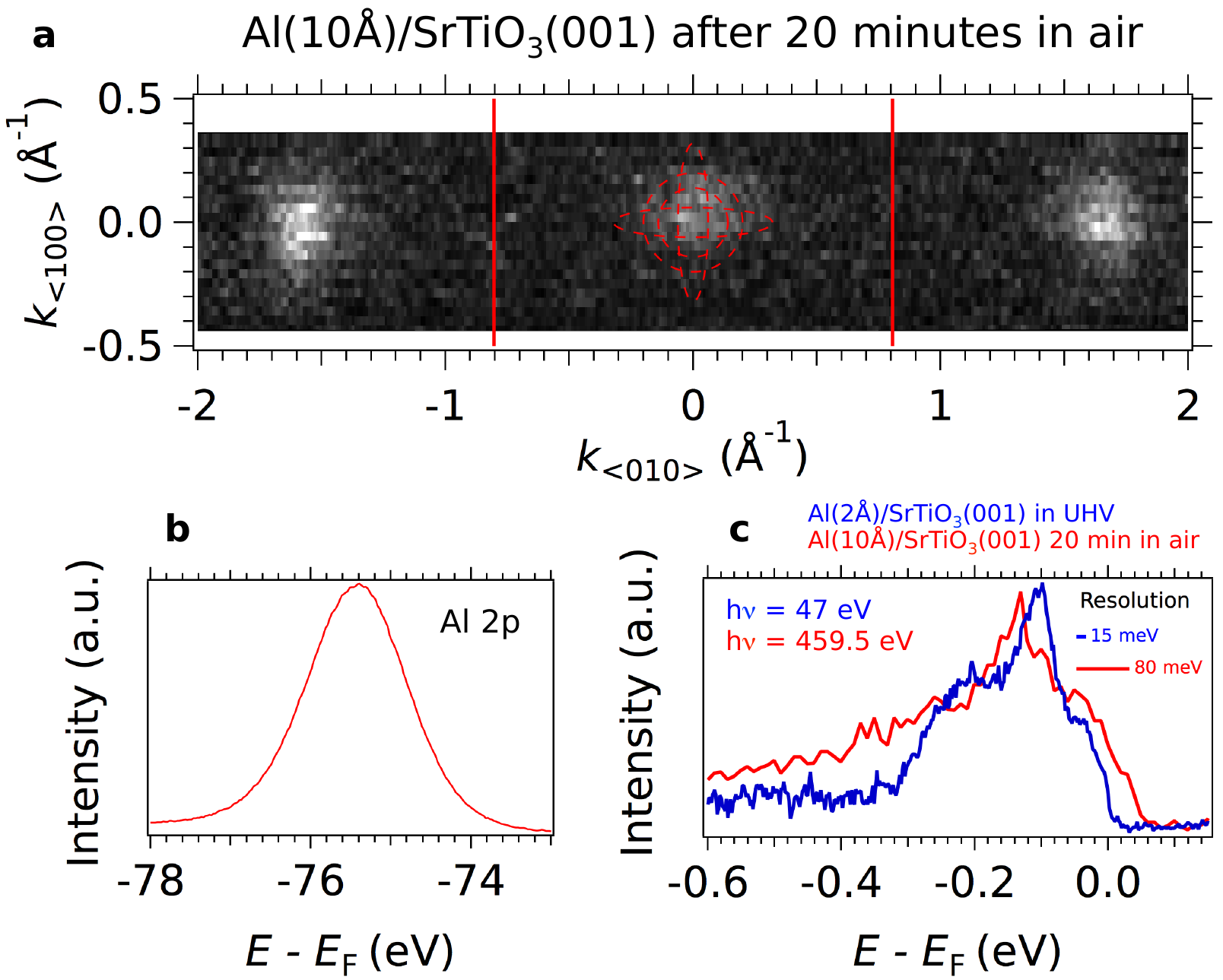}
    \end{center}
    \caption{\label{fig:high_hv} \footnotesize{
        (a) Fermi surface map measured in three neighboring Brillioun zones 
        at $h\nu=459.5$~eV on the Al(10\AA)/SrTiO$_3$(001) 
        interface prepared \emph{in-situ} and subsequently exposed to air 
        for $t=20$~min. The red dashed circles and ellipses illustrate the Fermi surface 
        shown in Figures~1(c,~d) of the main text.
        The thick red lines correspond to the borders of the bulk Brillioun zones. 
		(b) Angle-integrated spectrum of the Al 2p peak of the Al(10\AA)/SrTiO$_3$(001) 
		interface measured at $h\nu=458.4$~eV. 
		Due to the exposure to air, the Al film is completely oxidized 
		--compare to Figure~2 of the main text.
        (c) Energy distribution curves 
        integrated around $\Gamma$ measured at the Al(10\AA)/SrTiO$_3$(001) 
        and Al(2\AA)/SrTiO$_3$(001) interfaces.
        To facilitate the comparison, a momentum-independent background,
        due to spectral weight from the in-gap state, was removed
        from the EDC of the Al(2\AA)/SrTiO$_3$(001) data.
        }
      }     
\end{figure*} 

In principle, the thicker the oxidized Al film the better the passivation 
of the surface against re-oxidation in ambient air pressure. 
For amorphous Al$_2$O$_3$ films grown by atomic layer deposition on the surface of SrTiO$_3$(001), 
a film thickness of $\geq 1.2$~nm is sufficient to create a stable 2DES
at the Al$_2$O$_3$/SrTiO$_3$ interface~\cite{Lee2012}. 
Note that this value is identical to the thickness of the natural oxidized layer 
at the surface of aluminum ($1.24$~nm)~\cite{Cai2011}.
Hence, this thickness is sufficient to prevent oxygen diffusion 
through a homogenous Al$_2$O$_3$ capping layer.\\ 
In our case, the probing depth of the high-resolution ARPES measurements, 
such as the ones shown in the main text,
is limited by the mean free path of electrons which is $\sim 5$~\AA~at kinetic energies 
of $E_{kin}=20-100$~eV.
To increase the probing depth and probe the 2DES at buried interfaces, 
\emph{e.g.} LaAlO$_3$/SrTiO$_3$, soft x-ray angle-resolved resonant photoelectron spectroscopy 
was applied previously~\cite{Berner2013}.
Thus, to test the stability of the 2DES, we exposed an Al(10~\AA)/SrTiO$_3$ sample 
to ambient conditions for about 30 minutes, 
and conducted soft x-ray resonant ARPES ($hv$=459.5~eV) at low temperatures.
Note that $1$~nm of Al corresponds to about $1.25$~nm of Al$_2$O$_3$ 
which is close to the ``critical'' passivation thickness mentioned above.

As can be seen from the Fermi surface in \textbf{Figure~}\ref{fig:high_hv}(a), the 2DES 
at the Al(10~\AA)/SrTiO$_3$ interface
still exists after the exposure to air. 
For comparison,
the red dashed circles and ellipses represent the Fermi surfaces 
measured at the ultra-thin Al(2~\AA)/SrTiO$_3$ interface 
--see Figures~1(c,~d) of the main text. 
Note that the 10~\AA~Al layer was completely oxidized 
after exposure to air, as demonstrated by the peak shape of the Al-2p peak 
in Figure~\ref{fig:high_hv}(b). 
The data quality in these soft-X-ray ARPES measurements 
is lower compared to UV-measurements 
as the surface is not pristine anymore after exposure to air,
the 2DES is buried below a thick oxidized Al film,
the photoemission cross section of the valence states is much smaller 
at higher photon energies, and the total energy resolution 
at $hv = 459.5$~eV is about 80~meV, compared to 15~meV at $h\nu=47$~eV.
However, it is clear that the Fermi surface, and hence the charge-carrier density,
are comparable between the Al(10\AA)/SrTiO$_3$ sample exposed to air
and the pristine Al(2\AA)/SrTiO$_3$.
Figure~\ref{fig:high_hv}(c) compares the momentum-integrated band structure around $\Gamma$ 
for the two different interfaces, 
confirming that their electronic structures are comparable.
Thus, these results demonstrate that the oxidized Al/SrTiO$_3$ interface
effectively passivates the 2DES on SrTiO$_3$.  

In order to adapt the method of creation of 2DES at the Al/oxide interface to 
transport measurements, and to be certain that the oxidized Al layer completely 
blocks oxygen diffusion, a capping layer thickness above the ``critical'' passivation value
of $\sim 1.2$ nm is necessary. 
At the same time, the capping layer suitable for transport should be insulating 
without contributions of metallic Al.
Several possibilities should be explored in future studies:
optimization of growth parameters (\emph{e.g.} applying an oxygen partial pressure~\cite{Cai2011} 
during deposition after the first 2~\AA~of Al and/or a slight increase of the temperature 
to oxidize Al thicknesses greater than 2~\AA) 
or deposition of another type of insulating capping layer after the deposition of 2~\AA~of Al.

\subsection*{Subband dispersions at the Al(2\AA)/anatase interface}

\begin{figure}
    \begin{center}
        \includegraphics[clip, width=8cm]{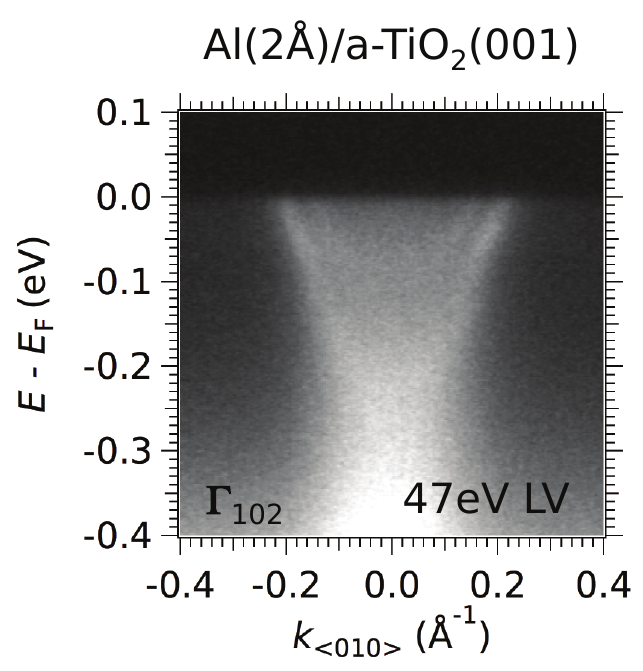}
    \end{center}
    \caption{\label{fig:Al_TiO2_001_EK} \footnotesize{
        Energy-momentum intensity map measured at the Al(2~\AA)/a-TiO$_2$(001) 
        interface prepared \emph{in-situ}.
        The data was recorded at $T=8$~K
        using LV photons at $h\nu = 47$~eV around the $\Gamma_{102}$ point.
        }
      }     
\end{figure}

\textbf{Figure~}\ref{fig:Al_TiO2_001_EK} presents the ARPES energy-momentum intensity map
at the Al(2\AA)/a-TiO$_2$(001) interface. The two subbands form the circular
Fermi surfaces shown in Figure~1(f) of the main text. 
As mentioned there, these Fermi surfaces are almost twice larger
than their counterparts at the surface of anatase reduced by photons~\cite{Roedel2015}.
In agreement with this observation, the two subbands
at the Al(2\AA)/a-TiO$_2$(001) interface disperse down to larger binding energies:
approximately $-100$~meV and $-230$~meV for the upper and lower subbands, respectively,
while at the bare, reduced anatase surface they disperse only down to about 
$-60$~meV and $-170$~meV~\cite{Roedel2015}. 

\subsection*{Oxygen vacancy distribution at the Al(2\AA)/anatase interface}
\begin{figure*}
  \begin{center}
   	  \includegraphics[clip, width=16cm]{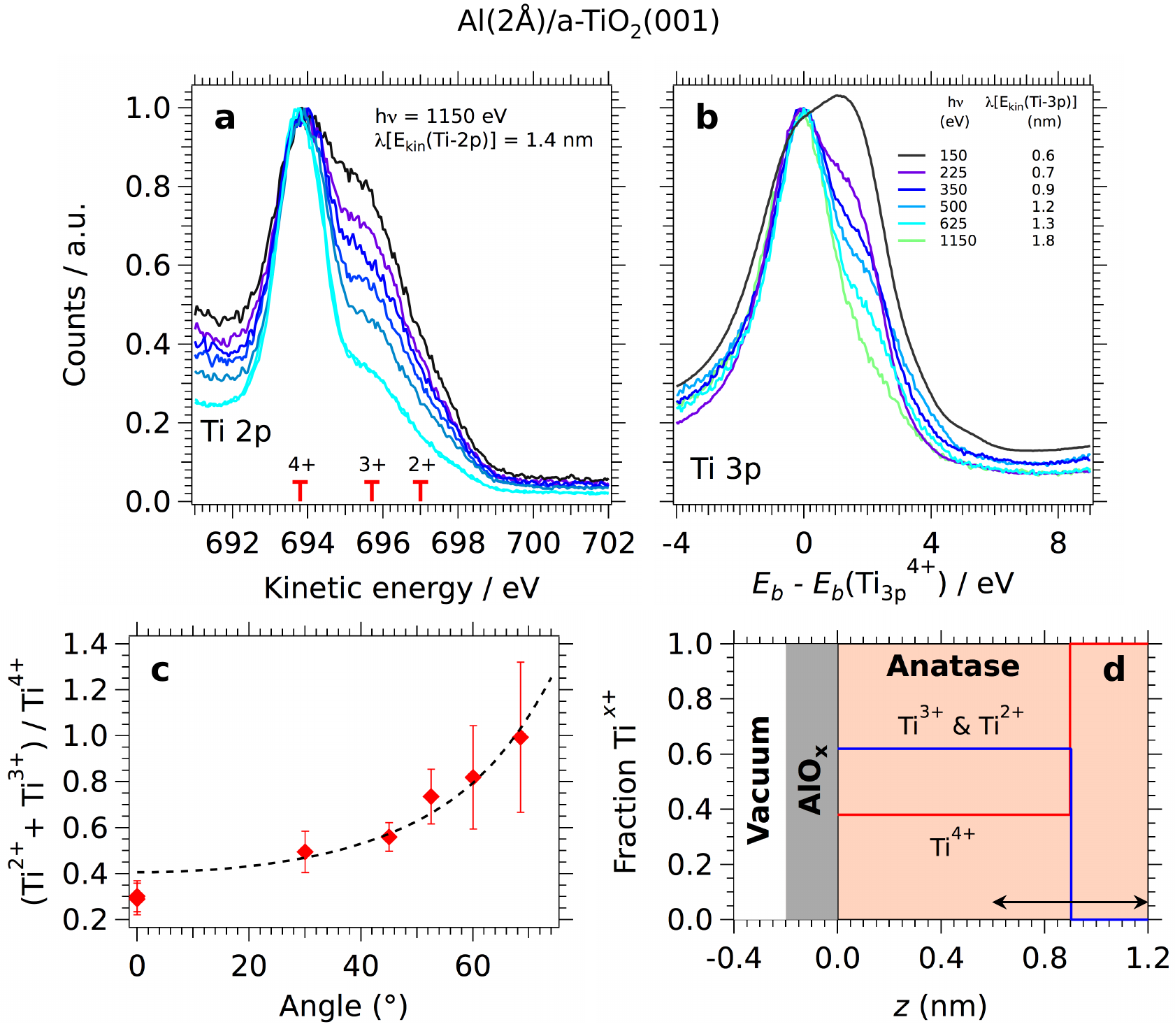}
  \end{center}
  \caption{\label{fig:TiO2_XPS_Ovacs} \footnotesize{
    (a)	XPS at the Ti-$2p$ core level of Al(2\AA)/a-TiO$_2$~$(001)$ as a function of emission angle
    	using $h\nu = 1150$~eV photons. At this photon energy, the universal 
    	inelastic mean free path of electrons emitted from the Ti-$2p$ peak
    	is $\lambda[E_{kin}(\textrm{Ti-}2p)] = 1.4$~nm~\cite{Seah1979}.
  	  	The red markers and corresponding error bars indicate the peak positions and uncertainties 
  	  	for the different Ti oxidation state ($4+$, $3+$, and $2+$).
  	(b) XPS at the Ti-$3p$ core level of anatase $(001)$ at normal emission 
  		as a function of photon energy. The inelastic mean free path of electrons
  		emitted from the Ti-$3p$ peak at different photon energies
  		is specified in the inset table.
  		The XPS intensity in panels (a) and (b) is normalized to the Ti$^{4+}$ peak.
  	(c)	Ratio of intensities, from panel (a), between the Ti$^{2+}$~$+$~Ti$^{3+}$ shoulder 
  		and the Ti$^{4+}$ peak as a function of the electron ejection angle. 
  		The dashed curve is the best fit to the data assuming a step-like distribution of vacancies
  		over $9 \pm 3$~\AA~below the surface, as schematized in panel~(d).
  	(d)	Model used for the distribution of the different Ti oxidation states
  		due to oxygen vacancies beneath the AlO$_x$/anatase interface:
  		blue line for Ti$^{2+}$~$+$~Ti$^{3+}$, red line for Ti$^{4+}$. The double arrow
  		indicates the error bar in the determination of the vacancy depth distribution.
  		}
  	} 
\end{figure*}

The redox reaction creates oxygen vacancies at the Al(2\AA)/oxide interface. 
The spatial distribution of these electron donors results in the creation of a potential well 
confining the electrons and forming the 2DES. 
To determine the distribution of vacancies,
we measured the Ti-$2p$ and Ti-$3p$ core levels of anatase $(001)$ 
using X-ray photoemission at $h\nu = 1150$~eV as a function of the electron emission angle, 
and at normal emission as a function of the X-ray photon energy,
and fitted the peaks using either Voigt or Lorentzian line shapes together with a Shirley background. 
As can be seen in \textbf{Figures~S}\ref{fig:TiO2_XPS_Ovacs}(a,~b), 
the core levels are composed of several peaks (red markers)
corresponding to Ti ions of different oxidation state ($4+$, $3+$, and $2+$) 
due to the presence of oxygen vacancies. 
We observe that the fraction of Ti-$4^{+}$ of stoichiometric, insulating TiO$_2$ 
increases for larger electron escape depths, as evidenced 
by the angle and photon energy dependencies in Figures~S\ref{fig:TiO2_XPS_Ovacs}(a,~b).
By contrats, the Ti-$3^{+}$ and Ti-$2^{+}$ components, 
associated to free carriers and oxygen vacancies, 
become increasingly important for smaller escape depths and thus, closer to the interface.\\
To obtain the concentration profile $c(z, Ti^{x+})$ of the $Ti^{x+}$ species 
along the confinement direction $z$ perpendicular to the surface, 
we calculate the total area of the corresponding core level peak by:
\begin{align*}
 a(Ti) \propto \int dz \sum_{x=2,3,4} c(z, Ti^{x+}) \exp\left(-\frac{d(z)}{\lambda(E_{kin})}\right),
\end{align*}  
where $d(z)$ is the distance travelled by a photo-emitted electron inside matter
(\emph{i.e.}, anatase~$+$~AlO$_x$ layer), which depends on the emission angle, 
and $\lambda$ the inelastic mean free path for electrons photo-emitted with kinetic energy $E_{kin}$. \\
Figure~\ref{fig:TiO2_XPS_Ovacs}(c) shows the ratio $[a(Ti^{2+})+a(Ti^{3+})]/a(Ti^{4+})$ 
as a function of the electron emission angle. 
The error bars indicate the variation of this ratio 
using different line shapes and backgrounds to fit the various Ti peaks. 
We then fit the observed changes in such peak area ratio using a Heavyside function 
for the concentration profile of oxygen vacancies, as shown in Figure~\ref{fig:TiO2_XPS_Ovacs}(d).
The result of the fit, shown by the dashed curve in Figure~\ref{fig:TiO2_XPS_Ovacs}(c),
yields a depth of $9 \pm 3$~\AA~for the vacancy-rich layer below the surface, 
and a fraction of $(62 \pm 15)$\% of Ti ions with oxidation states $3+$ or $2+$.

\subsection*{In-plane and out-of-plane Fermi surfaces of the 2DES in BaTiO$_3$}

\begin{figure*}
    \begin{center}
        \includegraphics[clip, width=16cm]{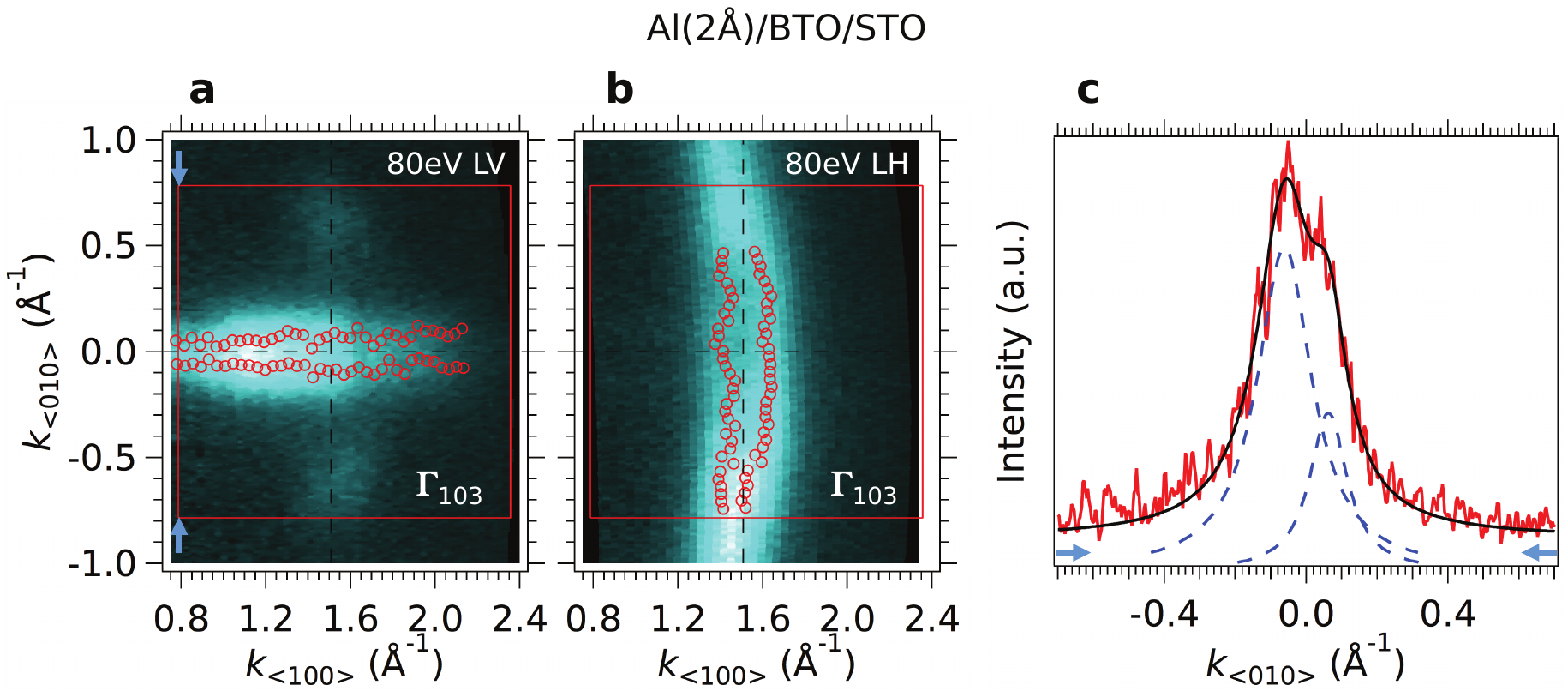}
    \end{center}
    \caption{\label{fig:BTO_KFs_Heavy} \footnotesize{
        (a,~b)~Fermi surface maps (spectral weight integrated over $E_F \pm 5$~meV)
        at the the Al(2\AA)/BaTiO$_3$ interface 
        using $80$~eV LV and LH photons, respectively. 
        Data were collected around the $\Gamma_{103}$ Brillouin zone.
        The open red circles show the Fermi momenta obtained from Lorentzian fits 
        to the MDCs at $E_F$. The red squares show the Brillouin-zone edges.
        (c)~MDC integrated over $E_F \pm 10$~meV 
        along the left edge of the $\Gamma_{103}$ Brillouin zone,
        corresponding to a cut along the light blue arrows in panel (a). 
        The blue dashed curves are Lorentzian peaks, and the black curve
        is the resulting total fit.    
        }
      }     
\end{figure*}

\textbf{Figures~S}\ref{fig:BTO_KFs_Heavy}(a,~b) show the Fermi-surface strips formed by the heavy bands 
of the 2DES at the surface of BaTiO$_3$ (Al(2\AA)/BTO/Nb:STO interface). 
The data were taken on the same Brillouin zone using mutually orthogonal photon polarizations,
which due to photoemission selection rules enhance either the Fermi strip parallel
to $k_{<010>}$ or the Fermi strip parallel to $k_{<100>}$.  
The open red circles show the Fermi momenta obtained from Lorentzian fits to the MDCs
at $E_F$.  Figure~\ref{fig:BTO_KFs_Heavy}(c) shows one of such MDCs, 
corresponding to a cut at the left edge of the 
$\Gamma_{103}$ Brillouin zone (light blue arrows).  
This MDC clearly shows a double-peak structure, corresponding to the two Fermi sheets 
of the Fermi strip, which is thus open at the Brillouin-zone edge. 
This is in agreement with the fact that the heavy bands running along the 
$k_{<010>}$ and $k_{<010>}$ directions do not cross $E_F$, 
as shown in Figures~3(d,~e) of the main text.
From these MDC fits, the average distance between opposite Fermi momenta 
along the short side of the Fermi strips is $k_{s} \approx 0.15 \pm 0.02$~\AA$^{-1}$.

\begin{figure*}
    \begin{center}
        \includegraphics[clip, width=16cm]{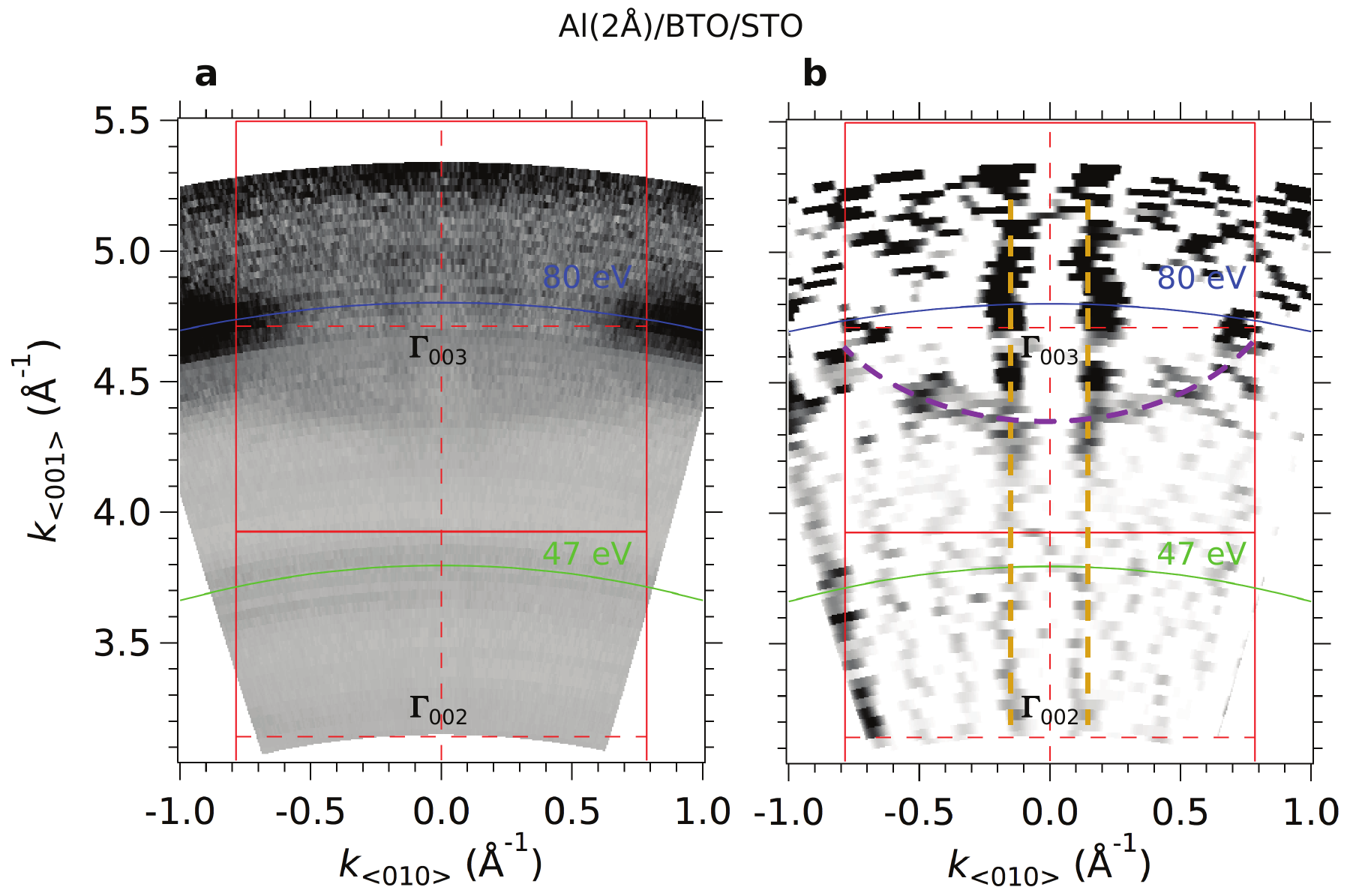}
    \end{center}
    \caption{\label{fig:BTO_hv_dep_d2} \footnotesize{
        (a)~Raw Fermi surface map at the the Al(2\AA)/BaTiO$_3$ interface 
		in the $k_{\langle 001 \rangle}$~--~$k_{\langle 010 \rangle}$ plane, 
		acquired by varying the photon energy in 1~eV steps 
		between $h\nu_1 = 30$~eV and $h\nu_2 = 100$~eV using LH photons. 
		To calculate the momentum perpendicular to the
		surface, we use a free-electron final-state approximation, 
		and set the inner potential to $V_0=12$~eV.
		The spectral weight was integrated over $[E_F - 30, E_F + 5]$~meV.
        The red square shows the edges of the $\Gamma_{003}$ bulk Brillouin zone.         
        The blue and green arcs show the spherical-cap cuts 
		in 3D reciprocal space obtained with $h\nu = 80$~eV and $h\nu = 47$~eV, 
		respectively, corresponding to the data presented in Figure~3 of the main text.
        (b)~Second derivative (negative values only) of the Fermi surface map in (a). 
        The yellow and purple dashed curves are guides to the eye showing, respectively,
        the non-dispersive Fermi surface of the light $d_{xy}$-like states,
        and the dispersive Fermi surface of the heavy $d_{xz/yz}$-like states. 
        }
      }     
\end{figure*}

Finally, \textbf{Figures~S}\ref{fig:BTO_hv_dep_d2}(a,~b) show the out-of-plane Fermi-surface map
of the 2DES at the surface of BaTiO$_3$,
obtained from the photon-energy dependence of the electronic structure 
measured over more than an entire bulk Brillouin zone.
The inner cylinder, yellow dashed lines in Figure~\ref{fig:BTO_hv_dep_d2}(b), 
is associated to the light $d_{xy}$-like band
forming the Fermi circle in the plane.  As its Fermi surface does not disperse along
the confinement direction, it corresponds to a 2D-like state.
The data also show a large ellipse dispersing along $k_{<001>}$, 
hence presenting a 3D-like character, best seen in the lower part of the $\Gamma_{003}$
Brillouin zone, purple dashed lines in Figure~\ref{fig:BTO_hv_dep_d2}(b). 
This Fermi sheet is associated to the heavy bands forming the Fermi strips in the plane. 
Of course, such 3D-like behavior cannot correspond to a true bulk state, 
as the redox reaction occurs only at the interface region. 
Note also that the 3D carrier density resulting from such a state would be huge, 
comparable to that of good metals, while the bulk BTO film is still insulating. 
Instead, such 2D-3D dichotomy between different states forming the 2DES in BaTiO$_3$, 
also observed for the 2DES at the surface of SrTiO$_3$~\cite{Plumb2014},
can be qualitatively understood as arising from confinement itself.
In the bulk, by cubic symmetry, the $t_{2g}$ bands are expected to form 
3 identical mutually orthogonal Fermi surfaces similar to prolate ellipsoids
along the main crystallographic axes
--or open quasi-cylinders, when the band filling is such that the ellipsoids'
long axis extends beyond the zone boundary.
Confinement along $z$ will result, by Heisenberg's principle, 
in ``de-confinement'', or elongation, of the ellipsoids along $k_z$.
When the confinement length becomes $\lesssim a$ (one unit cell), 
the ellipsoid stretches over $k_z \gtrsim 2\pi/a$ (one Brillouin zone),
and the out-of-plane Fermi surface becomes a cylinder.

Thus, in the case of the 2DES at the surface of BaTiO$_3$, 
we see from Figure~\ref{fig:BTO_KFs_Heavy} 
that the Fermi-surface strip formed by $d_{xz/yz}$-like states 
has an in-plane Fermi momentum $k_F = k_s/2 = 0.075$~\AA$^{-1}$, 
while from Figure~\ref{fig:BTO_hv_dep_d2} its out-of-plane Fermi momentum 
is approximately 0.4~\AA$^{-1}$.
Hence, there is an elongation along $k_z$ of the $d_{xz/yz}$ ellipsoids
due to confinement.
Similarly, as noted before, the in-plane circular Fermi surface, 
formed by $d_{xy}$-like states, forms a cylinder along the out-of-plane direction.
We then conclude that the planar $d_{xy}$-like states are more tightly confined
to the surface, while the non-planar $d_{xz/yz}$-like states extend over multiple
unit cells towards the bulk. This situation is wholly similar to the case
of the 2DES at the surface of SrTiO$_3$~\cite{Plumb2014}, and simply reflects the
fact that the confinement potential is wedge-shaped,
such that electrons with a large effective mass along $k_z$ ($d_{xy}$ states)
are more confined than electrons with a small effective mass along $k_z$ 
($d_{xz/yz}$ states)~\cite{Santander-Syro2011,Plumb2014}.


\pagebreak
   
\end{document}